\input harvmac.tex

%\draftmode

\def\ndt{\noindent}

\def\K3{{\bf K3}}
\def\journal#1&#2(#3){\unskip, \sl #1\ \bf #2 \rm(19#3) }
\def\andjournal#1&#2(#3){\sl #1~\bf #2 \rm (19#3) }

\def\bar{\overline}

\def\tilde{\widetilde}

\def\frac#1#2{{#1\over#2}}

\def\vev#1{\langle#1\rangle}
\def\d{\partial}

\def\inbar{\,\vrule height1.5ex width.4pt depth0pt}
\def\IC{\relax\hbox{$\inbar\kern-.3em{\rm C}$}}
\def\IR{\relax{\rm I\kern-.18em R}}
\def\IP{\relax{\rm I\kern-.18em P}}
\def\IN{\relax{\rm I\kern-.18em N}}

%
%%%%%%%%%%%%%%%%%%%%%%%%%%%%%%%%%%%%
%

%\def\ap#1#2#3{Ann. Phys. {\bf #1} (#2) #3}

%
\catcode`\@=11
\def\slash#1{\mathord{\mathpalette\c@ncel{#1}}}
\overfullrule=0pt

\def\LL{{\cal L}}
\def\NN{{\cal N}}
\def\OO{{\cal O}}

\def\lam{\lambda}

\def\underrel#1\over#2{\mathrel{\mathop{\kern\z@#1}\limits_{#2}}}

\catcode`\@=12

%%%%%%%%%%%%%%%%%%%%%%%%%%%%%%%%%%%%%%%%%%%%%%%%%%%%%%%%%%%%%%

%

\def\vev#1{\left\langle #1 \right\rangle}

%%%%%%%%%%%%%%%%%%%%%%%%%%%%%%%%%%%%%%%%%%%%%%%%%%%%%%%%%%%%%%
% new defs:

\def \ov {\over}
\def \p {\partial}
\def \ha {{1 \ov 2}}
\def \al {\alpha}
\def \lam {\lambda}

\def \om {\omega}
\def \Om {\Omega}
\def \ep {\epsilon}

\def\le{\left}
\def\ri{\right}

%%%%dotted indices

\def\IL{\relax{\rm I\kern-.18em L}}
\def\IH{\relax{\rm I\kern-.18em H}}
\def\IR{\relax{\rm I\kern-.18em R}}
\def\IC{\relax\hbox{$\inbar\kern-.3em{\rm C}$}}

%\def\IZ{{\bf Z}}
%\def\IR{{\bf R}}

%%%%%%%%%%%%%%%%%%%%%%%%%%%%

%%%% you need these macros:
%%%%%
%%%%%

%%% MACROS FOR BOX BOUNDARY CONDS
%%% FROM KAWAI ET AL

\def\makeblankbox#1#2{\hbox{\lower\dp0\vbox{\hidehrule{#1}{#2}%
   \kern -#1% overlap rules
   \hbox to \wd0{\hidevrule{#1}{#2}%
      \raise\ht0\vbox to #1{}% vrule height
      \lower\dp0\vtop to #1{}% vrule depth
      \hfil\hidevrule{#2}{#1}}%
   \kern-#1\hidehrule{#2}{#1}}}%
}%
\def\hidehrule#1#2{\kern-#1\hrule height#1 depth#2 \kern-#2}%
\def\hidevrule#1#2{\kern-#1{\dimen0=#1\advance\dimen0 by #2\vrule
    width\dimen0}\kern-#2}%
\def\openbox{\ht0=1.2mm \dp0=1.2mm \wd0=2.4mm  \raise 2.75pt
\makeblankbox {.25pt} {.25pt}  }

\def\bun#1/#2{\leavevmode
   \kern.1em \raise .5ex \hbox{\the\scriptfont0 #1}%
   \kern-.1em $/$%
   \kern-.15em \lower .25ex \hbox{\the\scriptfont0 #2}%
}

\def\opensquare{\ht0=3.4mm \dp0=3.4mm \wd0=6.8mm  \raise 2.7pt
\makeblankbox {.25pt} {.25pt}  }

%%%%%%%%%%%%%%%%%%%%%%%

\def\sector#1#2{\ {\scriptstyle #1}\hskip 1mm
\mathop{\opensquare}\limits_{\lower
1mm\hbox{$\scriptstyle#2$}}\hskip 1mm}

\def\tsector#1#2{\ {\scriptstyle #1}\hskip 1mm
\mathop{\opensquare}\limits_{\lower
1mm\hbox{$\scriptstyle#2$}}^\sim\hskip 1mm}
%%%
%%%

%%%%%%%%%%%%%%%%%%%%
%\Minw
%\SundborgUE
\lref\sundB{
  B.~Sundborg,
  ``The Hagedorn transition, deconfinement and N = 4 SYM theory,''
  Nucl.\ Phys.\ B {\bf 573}, 349 (2000)
  [arXiv:hep-th/9908001].
  %%CITATION = HEP-TH 9908001;%%
}

 \lref\MinW{
  O.~Aharony, J.~Marsano, S.~Minwalla, K.~Papadodimas and M.~Van Raamsdonk,
  ``The Hagedorn / deconfinement phase transition in weakly coupled large N
  gauge theories,''
  Adv.\ Theor.\ Math.\ Phys.\  {\bf 8}, 603 (2004)
  [arXiv:hep-th/0310285].
  %%CITATION = HEP-TH 0310285;%%
}

%\AharonyBQ
\lref\Minwi{
  O.~Aharony, J.~Marsano, S.~Minwalla, K.~Papadodimas and M.~Van Raamsdonk,
  ``A first order deconfinement transition in large N Yang-Mills theory on a
  small S**3,''
  Phys.\ Rev.\ D {\bf 71}, 125018 (2005)
  [arXiv:hep-th/0502149].
  %%CITATION = HEP-TH 0502149;%%
}

%\FuruuchiQM
\lref\FuruuchiQM{
  K.~Furuuchi,
  ``From free fields to AdS: Thermal case,''
  arXiv:hep-th/0505148.
  %%CITATION = HEP-TH 0505148;%%
}

%\MaldacenaRE
\lref\Malda{
  J.~M.~Maldacena,
  ``The large N limit of superconformal field theories and supergravity,''
  Adv.\ Theor.\ Math.\ Phys.\  {\bf 2}, 231 (1998)
  [Int.\ J.\ Theor.\ Phys.\  {\bf 38}, 1113 (1999)]
  [arXiv:hep-th/9711200].
  %%CITATION = HEP-TH 9711200;%%
}

%\GubserBC
\lref\gkp{
  S.~S.~Gubser, I.~R.~Klebanov and A.~M.~Polyakov,
  ``Gauge theory correlators from non-critical string theory,''
  Phys.\ Lett.\ B {\bf 428}, 105 (1998)
  [arXiv:hep-th/9802109].
  %%CITATION = HEP-TH 9802109;%%
}

%\WittenQJ
\lref\witt{
  E.~Witten,
  ``Anti-de Sitter space and holography,''
  Adv.\ Theor.\ Math.\ Phys.\  {\bf 2}, 253 (1998)
  [arXiv:hep-th/9802150].
  %%CITATION = HEP-TH 9802150;%%
}
%\WittenZW
\lref\wittM{
  E.~Witten,
  ``Anti-de Sitter space, thermal phase transition, and confinement in  gauge
  theories,''
  Adv.\ Theor.\ Math.\ Phys.\  {\bf 2}, 505 (1998)
  [arXiv:hep-th/9803131].
  %%CITATION = HEP-TH 9803131;%%
}
%\LeeBX
\lref\MinSe{
  S.~M.~Lee, S.~Minwalla, M.~Rangamani and N.~Seiberg,
  ``Three-point functions of chiral operators in D = 4, N = 4 SYM at  large
  N,''
  Adv.\ Theor.\ Math.\ Phys.\  {\bf 2}, 697 (1998)
  [arXiv:hep-th/9806074].
  %%CITATION = HEP-TH 9806074;%%
}

%\D'HokerEA
\lref\freedman{
  E.~D'Hoker, D.~Z.~Freedman, S.~D.~Mathur, A.~Matusis and L.~Rastelli,
  ``Extremal correlators in the AdS/CFT correspondence,''
  arXiv:hep-th/9908160.
  %%CITATION = HEP-TH 9908160;%%
}

%\PolchinskiTW
\lref\polchin{
  J.~Polchinski,
  ``High temperature limit of the confining phase,''
  Phys.\ Rev.\ Lett.\  {\bf 68}, 1267 (1992)
  [arXiv:hep-th/9109007].
  %%CITATION = HEP-TH 9109007;%%
}

%\IntriligatorFF
\lref\IntriligatorFF{
  K.~A.~Intriligator and W.~Skiba,
  ``Bonus symmetry and the operator product expansion of N = 4
  %super-Yang-Mills,''
  Nucl.\ Phys.\ B {\bf 559}, 165 (1999)
  [arXiv:hep-th/9905020].
  %%CITATION = HEP-TH 9905020;%%
}

%\GonzalezReyIH
\lref\Parketal{
  F.~Gonzalez-Rey, B.~Kulik and I.~Y.~Park,
  %``Non-renormalization of two point and three point correlators of N = 4  SYM
  %in N = 1 superspace,''
  Phys.\ Lett.\ B {\bf 455}, 164 (1999)
  [arXiv:hep-th/9903094].
  %%CITATION = HEP-TH 9903094;%%
}

%\HoweZI
\lref\HoweZI{
  P.~S.~Howe, E.~Sokatchev and P.~C.~West,
  %``3-point functions in N = 4 Yang-Mills,''
  Phys.\ Lett.\ B {\bf 444}, 341 (1998)
  [arXiv:hep-th/9808162].
  %%CITATION = HEP-TH 9808162;%%
}

%\EdenGH
\lref\EdenGH{
  B.~Eden, P.~S.~Howe and P.~C.~West,
  ``Nilpotent invariants in N = 4 SYM,''
  Phys.\ Lett.\ B {\bf 463}, 19 (1999)
  [arXiv:hep-th/9905085].
  %%CITATION = HEP-TH 9905085;
}

%\D'HokerTZ
\lref\DHokerTZ{
  E.~D'Hoker, D.~Z.~Freedman and W.~Skiba,
  ``Field theory tests for correlators in the AdS/CFT correspondence,''
  Phys.\ Rev.\ D {\bf 59}, 045008 (1999)
  [arXiv:hep-th/9807098]; \quad
  %%CITATION = HEP-TH 9807098;%%
}

\lref\MetsaevIT{
  R.~R.~Metsaev and A.~A.~Tseytlin,
  ``Type IIB superstring action in AdS(5) x S(5) background,''
  Nucl.\ Phys.\ B {\bf 533}, 109 (1998)
  [arXiv:hep-th/9805028].
  %%CITATION = HEP-TH 9805028;%%
}

%\DHokerTZ\IntriligatorFF\EdenGH

%\AharonyIG
\lref\AharonyIG{
  O.~Aharony, J.~Marsano, S.~Minwalla and T.~Wiseman,
  ``Black hole - black string phase transitions in thermal 1+1 dimensional
  supersymmetric Yang-Mills theory on a circle,''
  Class.\ Quant.\ Grav.\  {\bf 21}, 5169 (2004)
  [arXiv:hep-th/0406210].
  %%CITATION = HEP-TH 0406210;%%
}

% \lref\Minw{hep-th0310285 , Phys.\ Rept.\  }
%\lref\NewMinw{hep-th/0502149} \lref\bigR{big review }
%\lref\hongguido{hong and guido, notes } \lref\notes{ym\_g.tex}
%\lref\book{  Fujii, Yasunori.The scalar-tensor theory of gravitation
%/Yasunori Fujii, Kei-ichi Maeda.  QC178.F85 2003 }
%\lref\unruh{Unruh, W. G. Unruh, "Alternative Fock quantization of
%neutrinos in flat space-time} \lref\freedmanmathur {Freedman,
%Mathur, Matusis and Rastelli}
%\CrnkovicMS
%\lref\CrnkovicMS{ C.~Crnkovic, M.~R.~Douglas and G.~W.~Moore,
%``Physical Solutions For Unitary Matrix Models,'' Nucl.\ Phys.\ B
%{\bf 360}, 507 (1991).
%%CITATION = NUPHA,B360,507;%%
%}

%\BerkovitsXU
\lref\berkovits{
  N.~Berkovits,
  ``Quantum consistency of the superstring in AdS(5) x S**5 background,''
  JHEP {\bf 0503}, 041 (2005)
  [arXiv:hep-th/0411170].
  %%CITATION = HEP-TH 0411170;%%
}

%\PolchinskiZF
\lref\PolchinskiZF{
  J.~Polchinski,
  ``Evaluation Of The One Loop String Path Integral,''
  Commun.\ Math.\ Phys.\  {\bf 104}, 37 (1986).
  %%CITATION = CMPHA,104,37;%%
}

%\DolanIY
\lref\osborn{
  F.~A.~Dolan and H.~Osborn,
  ``Conformal partial wave expansions for N = 4 chiral four point  functions,''
  arXiv:hep-th/0412335.
  %%CITATION = HEP-TH 0412335;%%
}

%\FuruuchiEU
\lref\FuruuchiEU{
  K.~Furuuchi,
  ``Large N reductions and holography,''
  arXiv:hep-th/0506183.
  %%CITATION = HEP-TH 0506183;%%
}

%\KalloshQS
\lref\kallosh{
  R.~Kallosh and A.~Rajaraman,
  ``Vacua of M-theory and string theory,''
  Phys.\ Rev.\ D {\bf 58}, 125003 (1998)
  [arXiv:hep-th/9805041].
  %%CITATION = HEP-TH 9805041;%%
}

%%%%%%%%%%%%%%%%%%%%%%%%%%%%%%

\Title{\vbox{\baselineskip12pt \hbox{hep-th/0509117}
\hbox{MIT-CTP-3679}
}}%
 {\vbox{\centerline{Inheritance principle and }
 \medskip
 \smallskip
 \centerline{ Non-renormalization theorems at finite temperature}
 %\smallskip
 %{Black hole singularities in
 %Yang-Mills
 %theories (I)
  }}

\smallskip
\centerline{Mauro Brigante$^1$, Guido Festuccia$^{1,2}$ and Hong
Liu$^{1,2}$ }
\medskip

\centerline{$^1$ {\it  Center for Theoretical Physics}}
\centerline{\it Massachusetts Institute of Technology}
\centerline{\it Cambridge, Massachusetts, 02139}
\smallskip
\centerline{$^2$ {\it School of Natural Sciences}} \centerline{\it
Institute for Advanced Study} \centerline{\it Einstein Drive,
Princeton, NJ 08540}

\smallskip

\vglue .3cm

\bigskip
\noindent

We present a general proof of an ``inheritance principle'' satisfied
by a weakly coupled $SU(N)$ gauge theory with adjoint matter on a
class of compact manifolds (like $S^3$). In the large $N$ limit,
finite temperature correlation functions of gauge invariant
single-trace operators in the low temperature phase are related to
those at zero temperature by summing over images of each operator in
the Euclidean time direction.  As a consequence, various
non-renormalization theorems of $\NN=4$ Super-Yang-Mills theory on
$S^3$ survive at finite temperature despite the fact that the
conformal and supersymmetries are both broken.

 \Date{Sep. 15, 2005}

%%%%%%%%%%%%%%%%%%%%%%%%%%%%%%%%%%%%%%%%%%%%%%%%%%%%%%

%\vfil\eject
\bigskip

%%%%%%%%%%%%%%%%%%%%%%%%%%%%

%\draftmode %\listtoc \writetoc

\newsec{Introduction}

It has been shown that weakly coupled $SU(N)$ gauge theories with
adjoint matter on a class of compact manifolds (including $S^3$)
have a large $N$  ``deconfinement'' transition at a temperature
$T_c$~\refs{\sundB,\MinW,\Minwi}. In the low temperature
(``confined'') phase $T< T_c$, the free energy is of order $O(1)$,
while in the high temperature (``deconfined'') phase the free energy
becomes of order $O(N^2)$.

The main purpose of the paper is to give a general proof of an
``inheritance principle'' satisfied by these gauge theories in the
low temperature phase and point out some consequences of it. More
explicitly, suppose at zero temperature the {\it Euclidean}
$n$-point function of some gauge invariant single trace operators
$\OO_1, \cdots \OO_n$ is given by
$$
G_0 (\tau_1,e_1; \tau_2,e_2; \cdots; \tau_n,e_n) = \vev{\OO_1
(\tau_1,e_1) \OO_2 (\tau_2,e_2) \cdots \OO_n (\tau_n,e_n)}
$$
where $\tau_i$ denote the Euclidean time and $e_i$ denote a point on
the compact manifold. Then one finds that the corresponding
correlation function at finite temperature $T = {1 \ov \beta}$ is
given by\foot{The following expression assumes that all $\OO_i$ are
bosonic. If an $\OO_i$ is fermionic, then one multiplies an
additional factor $(-1)^{m_i}$.}
 \eqn\sunG{
 G_\beta (\tau_1, \tau_2, \cdots,
\tau_n) = \sum_{m_1, \cdots m_n=-\infty}^\infty G_0 (\tau_1 - m_1
\beta,
 \tau_2 - m_2 \beta, \cdots , \tau_n - m_n \beta)
 }
where for notational simplicity we have suppressed the spatial
coordinates. In other words, one adds images for each operator
$\OO_i$ in the Euclidean time direction. Note that the statement is
not trivial, since in thermal gauge theory computations one is
supposed to add images for each fundamental field in the operator,
not the operator as a whole.

Equation~\sunG\ was first derived in~\refs{\FuruuchiQM}, where the
case of a scalar field on $\IR^3$ was explicitly considered. The
crucial ingredients for establishing \sunG\ were the dominance of
planar graphs in the large $N$ limit and the saddle point
configuration of the time component $A_0$ of the gauge field
characteristic of the confined phase. While not shown explicitly,
the discussion there could in principle be generalized to compact
spaces and including dynamical gauge fields and fermions.

In this paper we will present an alternative proof of \sunG\ for
gauge theories on compact spaces including dynamical gauge fields
and fermions. Our framework has the advantages that the discussions
can be easily generalized to the study of non-planar diagrams and
other saddle points of finite temperature gauge theories. We also
discuss some subtleties involving self-contractions at finite
temperature and present an explicit argument that \sunG\ holds to
{\it all orders} in perturbative expansions in the `t~Hooft
coupling.

Equation \sunG\ implies that properties of correlation functions of
the theory at zero temperature can be inherited at finite
temperature in the large $N$ limit. For example, for those
correlation functions which are independent of the 't Hooft coupling
in the large $N$ limit at zero temperature, the statement remains
true at finite temperature. For $\NN=4$ Super-Yang-Mills theory
(SYM) on $S^3$, which was the main motivation of our study, it was
conjectured in~\refs{\MinSe} that two- and three-point functions of
chiral operators are nonrenormalized from weak to strong
coupling\foot{See
also~\refs{\DHokerTZ,\IntriligatorFF,\EdenGH,\Parketal} for further
evidence.}. The conjecture, if true, will also hold for $\NN=4$ SYM
theory at finite temperature despite the fact that the conformal and
supersymmetries are broken. \sunG\ also suggests that at leading
order in $1/N$ expansion, the one-point functions of all gauge
invariant operators (including the stress tensor) at finite
temperature are zero.

While \sunG\ is somewhat non-intuitive from the gauge theory point
of view, it has a simple interpretation\foot{It also has a natural
interpretation from the point of view of large $N$ reduction. This
was pointed out to us by S. Shenker and K. Furuuchi. See
also~\refs{\FuruuchiEU }.} from the string theory
dual~\refs{\FuruuchiQM}. Suppose that the gauge theory under
consideration is described by a string theory on some target space
$M$. Then \sunG\ translates into the statement that the theory at
finite temperature is described by propagating strings on $M$ with
Euclidean time direction compactified with a period $\beta$. The
leading order expression for $G_\beta$ in large $N$ limit is mapped
to the sphere amplitude of the dual string theory. The
``inheritance'' principle in the gauge theory then follows simply
from that of the tree level orbifold string theory.

We note that given a perturbative string theory, it is not {\it a
priori} obvious that the theory at finite temperature is described
by the same target space with time direction periodically
identified\foot{A counter example is IIB string in $AdS_5 \times
S_5$ at a temperature above the Hawking-Page temperature. Also in
curved spacetime this implies one has to choose a particular time
slicing of the spacetime.}. For perturbative string theory in flat
space at a temperature below the Hagedorn temperature, this can be
checked by explicit computation of the free energy at
one-loop~\refs{\PolchinskiZF}. Equation \sunG\ provides evidences
that this should be the case for string theories dual to the class
of gauge theories we are considering at a temperature $T<T_c$.

For $\NN=4$ SYM theory on $S^3$, the result matches well with that
from the AdS/CFT correspondence~\refs{\Malda,\gkp,\witt}. The
correspondence implies that when the curvature radius of the anti-de
Sitter (AdS) spacetime is much larger than the string and Planck
scales, which corresponds to the YM theory at large 't Hooft
coupling, IIB string in $AdS_5 \times S_5$ at $T <T_c$ is described
by compactifying the time direction (so-called thermal
AdS)~\refs{\witt,\wittM}. The result from the weakly coupled side
suggests that this description can be extrapolated to weak
coupling\foot{Also note that it is likely that $AdS_5 \times S_5$ is
an exact string background~\refs{\MetsaevIT,\kallosh,\berkovits}.}.

The plan of the paper is as follows. In section two we make some
general statements regarding finite temperature correlation
functions in free theory. In section three we prove the
``inheritance principle'' in free theory limit. In section four we
generalize the discussion to include interactions. In section five
we conclude with a discussion of string theory interpretation and
some other remarks.

For the rest of the paper unless stated explicitly, by finite
temperature we always refer to finite temperature in the low
temperature phase of the theory.

Note added in second version: We appreciated that \sunG\ had been
essentially derived in~\FuruuchiQM\ only after the submission of the
first version of this paper\foot{We thank K.~Furuuchi for
emphasizing this to us.}, in which~\FuruuchiQM\ was not given
sufficient recognition.

\newsec{Correlation functions of free Yang-Mills theory on $S^3$}

In this section we discuss some general aspects of free gauge
theories with adjoint matter on $S^3$ at finite temperature. We will
assume that the theory under consideration has a vector field
$A_\mu$ and a number of scalar and fermionic fields\foot{We also
assume that the scalar fields are conformally coupled.} all in the
adjoint representation of $SU(N)$. The discussion should also be
valid for other simply-connected compact manifolds. We use the
Euclidean time formalism with time direction $\tau$ compactified
with a period $\beta= {1 \ov T}$. Spacetime indices are denoted by
$\mu = (\tau, i)$ with $i$ along directions on $S^3$.

 The theory on $S^3$ can be
written as a $(0+1)$-dimensional (Euclidean) quantum mechanical
system by expanding all fields in terms of spherical harmonics on
$S^3$. Matter scalar and fermionic fields can be expanded in terms
of scalar and spinor harmonics respectively. For gauge field, it is
convenient to use the Coulomb gauge $\nabla_i A^i =0$, where
$\nabla$ denotes the covariant derivative on $S^3$. In this gauge,
$A_i$ can be expanded in terms of transverse vector harmonics,
$A_\tau$ and the Fadeev-Popov ghost $c$ can be expanded in terms of
scalar harmonics. At quadratic level, the resulting action has the
form
 \eqn\onedAc{\eqalign{
 S_{0} & = N \Tr \int\!d\tau  \;
\biggl[  \;  %\sum_{a}\,
 \le(\frac{1}{2} (D_\tau M_a)^2
 - \ha \om_{a}^2 M_a^2 \ri)  +  % \sum_{a}
 \xi_{a}^\dagger (D_\tau + \tilde \om_{a}) \xi_{a} + \ha m_{a}^2 v_a^2 +
 m_{a}^2 \bar c_a c_a
    \, \biggr] \  \cr
   %& \qquad + \ha \om_a^2 v_a^2 + \om_a^2 \bar c_a c_a \cr
 }}
where we have grouped all harmonic modes into three groups:

\item{1.} Bosonic modes $M_a$ with nontrivial kinetic terms. Note
that in the Coulomb gauge, the harmonic modes of the dynamical gauge
fields have the same $(0+1)$-d action as those from matter scalar
fields. We thus use $M_a$ to denote  collectively harmonic modes
coming from both the gauge field $A_i$ and matter scalar fields.

\item{2.} Fermionic modes $\xi_a$ with nontrivial kinetic terms.

\item{3.} $v_a$ and $c_a$ are from nonzero modes of $A_\tau$ and
the Fadeev-Popov ghost $c$, which have no kinetic terms.

\ndt The explicit expressions of various $(0+1)$-dimensional masses
$\om_{a}, \tilde \om_a, m_{a}$ can in principle be obtained from
properties of various spherical harmonics and will not be used
below. In \onedAc, following~\refs{\MinW} we separated the zero mode
$\al (\tau)$ of $A_\tau$ on $S^3$ from the higher harmonics and
combine it with $\p_\tau$ to form the covariant derivative $D_\tau$
of the $(0+1)$-dimensional theory, with
$$
D_\tau M_a= \p_\tau M_a - i [\al, M_a], \qquad
 D_\tau \xi_a= \p_\tau \xi_a - i [\al, \xi_a] \ .
$$
$\al (\tau)$ plays the role of the Lagrange multiplier which imposes
the Gauss law on physical states. In the free theory limit the ghost
modes $c_a$ do not play a role and $v_a$ only give rise to contact
terms (i.e. terms proportional to delta functions in the time
direction) in correlation functions\foot{Also note that since $v_a,
c_a$ do not have kinetic terms, at free theory level they only
contribute to the partition function by an irrelevant
temperature-independent overall factor.}. %We will thus ignore them
%in our free theory discussion of this and next sections.
Also note that $M_a, \xi_a$ satisfy periodic and anti-periodic
boundary conditions respectively
 \eqn\ndbo{
 M_a (\tau + \beta) = M_a (\tau), \qquad \xi_a (\tau + \beta) = -\xi_a
 (\tau)\ .
 }

Upon harmonic expansion, correlation functions of gauge invariant
operators in the four-dimensional theory reduce to sums of those of
the one-dimensional theory \onedAc. More explicitly, a
four-dimensional operator $\OO(\tau, e)$ can be expanded as
 \eqn\opred{
 \OO (\tau, e) = \sum_i f^{(\OO)}_i (e) Q_i (\tau)
 }
where $e$ denotes a point on $S^3$ and $Q_{i}$ are operators formed
from $M_a, \xi_a, v_a$ and their time derivatives. The functions
$f_{i}^{(\OO)} (e)$ are given by products of various spherical
harmonics. A generic $n$-point function in the four-dimensional
theory can be written as:
 \eqn\rysk{
 \vev{\OO_{1} (\tau_1,e_1) \OO_{2} (\tau_2,e_2) \cdots \OO_n
 (\tau_n,e_n)}
 = \sum_{i_1, \cdots, i_n} f_{i_1}^{\OO_1} (e_1) \cdots
 f_{i_n}^{\OO_n}(e_n) \,
 \vev{Q_{i_1} (\tau_1) Q_{i_2} (\tau_2) \cdots Q_{i_n}
 (\tau_n)}
 }
where $\vev{\cdots}$ on the right hand side denotes correlation
functions in the 1-dimensional theory \onedAc. Note that \rysk\
applies to all temperatures.

The theory \onedAc\ has a residue gauge symmetry
 \eqn\gauS{\eqalign{
 M_a &\to  \Om M_a \Om^\dagger , \qquad
 \xi_a  \to  \Om \xi_a \Om^\dagger \cr
 \al  & \to  \Om \al \Om^\dagger + {i} \Om \p_\tau \Om^\dagger \ . \cr
 }}
At zero temperature, the $\tau$ direction is uncompact. One can use
the gauge symmetry \gauS\ to set $\al=0$. Correlation functions of
the theory \onedAc\ can be obtained from the propagators of $M_a,
\xi_a$ by Wick contractions. Note that\foot{We use
$\vev{\cdots}_{0}$ and $\vev{\cdots}_{\beta}$ to denote the
correlation functions of \onedAc\ at zero and finite temperature
respectively.}
 \eqn\powa{\eqalign{
 & \vev{M_{ij}^a (\tau) \, M_{kl}^b (0)}_0
 =  {1 \ov N} G_s(\tau;\om_a) \delta_{ab} \delta_{il} \delta_{kj} \cr
 &  \vev{\xi_{ij}^a (\tau) \, \xi_{kl}^b (0)}_0
 = {1 \ov N} G_f(\tau;\tilde \om_a) \delta_{ab} \delta_{il} \delta_{kj} \cr
 }}
where
 \eqn\oenrP{
G_s (\tau; \om)  = {1 \ov 2 \om} e^{-\om|\tau|}, \qquad
 G_f(\tau; \om)  = (-\d_\tau +\om
)G_s(\tau;\om) \ .
 }
and $i,j,k,l$ denote $SU(N)$ indices.

At finite temperature, one can again use a gauge transformation to
set $\al (\tau)$ to zero. The gauge transformation, however,
modifies the boundary conditions from \ndbo\ to
 \eqn\nonTB{
 M_a(\tau + \beta) = U M_a U^\dagger ,
 \qquad \xi_a(\tau + \beta) = -U \xi_a U^\dagger \ .
 }
The unitary matrix $U$ can be understood as the Wilson line of $\al$
wound around the $\tau$ direction, which cannot be gauged away. It
follows that the path integral for \onedAc\ at finite $T$ can be
written as
 \eqn\pathI{
 \vev{\cdots}_{\beta} = {1 \ov Z(\beta)} \int dU  \int D M (\tau) D \xi(\tau) \,
 \cdots \; e^{-S_{0}[M_a, \xi_a; \al =0]}
 }
with $M_a, \xi_a$ satisfying boundary conditions \nonTB\ and $Z$ the
partition function.
 %Note that
%in \pathI\ the functional integral over $\al (\tau)$ is replaced
%by the matrix integral over $U$.

Since the action \onedAc\ has only quadratic dependence on $M_a$ and
$\xi_a$, the functional integrals over $M_a$ and $\xi_a$ in \pathI\
can be carried out straightforwardly, reducing \pathI\ to a matrix
integral over $U$. For example, the partition function can be
written as
 \eqn\parT{
 Z (\beta) = \int dU \; e^{S_{eff} (U)}
 }
where $S_{eff} (U)$ was computed in~\refs{\sundB,\MinW}
 \eqn\wisp{
 S_{eff} (U) = \sum_{n=1}^\infty {1 \ov n} V_n (\beta) \Tr U^n \Tr U^{-n}
 }
with
$$
V_n (\beta) = z_s (n\beta) +
 (-1)^{n+1}
 z_f (n \beta) , \qquad z_s (\beta) = \sum_a e^{-\beta \om_a}, \qquad z_f
(\beta) = \sum_a e^{-\beta \tilde \om_a} \ .
$$
Similarly, correlation functions at finite temperature are obtained
by first performing Wick contractions and then evaluating the matrix
integral for $U$. With boundary conditions \nonTB, the contractions
of $M_a$ and $\xi_a$ become
 \eqn\fincon{\eqalign{
 & \underbrace{M_{ij}^a (\tau) \, M_{kl}^b (0)}
 = {\delta_{ab} \ov N} \sum_{m=-\infty}^\infty G_s(\tau - m \beta; \om_a)
 U^{-m}_{il} U^{m}_{kj}
 \cr
 & \underbrace{\xi_{ij}^a (\tau) \, \xi_{kl}^b (0)}
 = {\delta_{ab} \ov N} \sum_{m=-\infty}^\infty (-1)^m G_f(\tau - m \beta; \tilde \om_a)
 U^{-m}_{il} U^{m}_{kj} \ .
 \cr
 }}
\fincon\ are obtained from \powa\ by summing over images in
$\tau$-direction and can be checked to satisfy \nonTB.

As an example, let us consider the planar expression of one- and
two-point functions of a normal-ordered operator $Q = \Tr M^4$, with
$M$ being one of the $M_a$ in \onedAc. One finds that
 \eqn\anexm{\eqalign{
 \vev{\Tr M^4}_{\beta} & = {2 \ov N^2} \sum_{m \neq 0, n \neq 0}
 G_s (-m \beta) G_s (-n \beta) \; \vev{\Tr U^m \Tr U^n \Tr U^{-m-n}}_U \cr
 }}
and the connected part of the two-point function is
 \eqn\twoPf{\eqalign{
  & \vev{\Tr M^4 (\tau) \Tr M^4 (0)}_{\beta} \cr
 & = {4 \ov N^4} \sum_{m,n,p,q} G_s (\tau - m\beta) G_s (\tau - n\beta)
  G_s (\tau - p\beta) G_s (\tau - q\beta) \; \vev{\Tr U^{q-m} \Tr U^{m-n} \Tr U^{n-p} \Tr
  U^{p-q}}_U \cr
  & + {16 \ov N^4} \sum_{m,n \neq 0, p,q} G_s (- m\beta) G_s (- n\beta)
  G_s (\tau - p\beta) G_s (\tau - q\beta) \cr
   & \qquad \times  \vev{\Tr U^{m} \Tr U^{n} \le(\Tr U^{-m-p+q} \Tr
  U^{-n+p-q} + \Tr U^{-m-p-n+q} \Tr
  U^{p-q} \ri)}_U \cr
 }}
In \anexm-\twoPf\ all sums are from $-\infty$ to $+\infty$ %$G_s$
%was given in \oenrP,
and
 \eqn\usna{
 \vev{\cdots}_U = {1 \ov Z} \int dU \; \cdots \;
e^{S_{eff} (U)}
 }
with $Z$ given by \parT. We conclude this section by noting some
features of \anexm-\twoPf:

\item{1.} Since the operators are normal-ordered, the zero
temperature contributions to the self-contractions (corresponding to
$m,n=0$) are not considered. In general, the one-point function is
not zero at finite $T$ because of the sum over images; this is clear
from \anexm.

\item{2.} The first term of \twoPf\ arises from contractions in
which all $M$'s of the first operator contract with those of the
second operator. The second term of \twoPf\ contains partial
self-contractions\foot{Full self-contractions correspond to
disconnected contributions.}, i.e. two of $M$'s in $\Tr M^4$
contract within the operator. The non-vanishing of self-contractions
is again due to the sum over nonzero images.

\newsec{Correlation functions in the low temperature phase}

It was found in~\refs{\sundB,\MinW} that \onedAc\ has a first order
phase transition at a temperature $T_c$ in the $N=\infty$ limit.
$\Tr U^n$ can be considered as order parameters of the phase
transition. In the low temperature phase, one has
 \eqn\Imeve{
 \vev{\Tr U^n}_U \approx N \delta_{n,0} + O(1/N)
 }
while for $T>T_c$, $\Tr U^n, n \neq 0$ develop nonzero expectation
values. It follows from \Imeve\ that in the low temperature phase,
to leading order in $1/N$ expansion
 \eqn\factE{\eqalign{
 & \vev{\Tr U^{n_1} \Tr U^{n_2} \cdots \Tr U^{n_k}}_U \cr
 & \approx
 \vev{\Tr U^{n_1}}_U \vev{ \Tr U^{n_2}}_U \cdots \vev{\Tr U^{n_k}}_U \cr
 & \approx N^k \delta_{n_1,0} \cdots \delta_{n_k, 0}
 }}
where in the second line we have used the standard factorization
property at large $N$.

We now look at the implications of \factE\ on correlation functions.
Applying \factE\ to \anexm\ and \twoPf, one finds
 \eqn\newExn{\eqalign{
 \vev{\Tr M^4}_{\beta} & = 0 + O(1/N)\cr
  \vev{\Tr M^4 (\tau) \Tr M^4 (0)}_{\beta} & =
  {4 } \sum_{m} G_s^4 (\tau - m\beta) + O(1/N^2) \cr
  & = \sum_{m} \vev{\Tr M^4 (\tau - m \beta) \Tr M^4 (0)}_{0}
 }}
Note that the second term of \twoPf\ due to partial
self-contractions vanishes and the finite-temperature correlators
are related to the zero-temperature ones by adding the images for
the whole operator.

The conclusion is not special to \newExn\ and can be generalized to
any correlation functions of single-trace (normal-ordered) operators
in the large $N$ limit. Now consider a generic $n$-point function
for some single-trace operators. At zero temperature, the
contribution of a typical contraction can be written in a form
 \eqn\zerTe{
 {1 \ov N^{n-2+2h}} \prod_{i<j=1}^n \prod_{p=1}^{I_{ij}}
 G^{(p)}_s(\tau_{ij}), \qquad \tau_{ij} = \tau_i - \tau_j
 }
where $i,j$ enumerate the vertices, $I_{ij}$ is the number of
propagators between vertices $i,j$, $G^{(p)} (\tau_{ij})$ is the
$p$-th propagator between vertices $i$ and $j$, and $h$ is the genus
of the diagram.  At finite temperature, one uses \fincon\ to add
images for each propagator and finds the contribution of the same
diagram is given by
 \eqn\genrTm{
 {1 \ov N^I} \le(\prod_{i < j=1}^n \prod_{p=1}^{I_{ij}}
  \sum_{m_{ij}^{(p)}=-\infty}^\infty \ri)
  \le(\prod_{i < j=1}^n \prod_{p=1}^{I_{ij}} %\sum_{m_{ij}^{(1)},m_{ij}^{(2)} \cdots}
 G_s^{(p)} \le(\tau_{ij} - m_{ij}^{(p)} \beta\ri) \ri)
 \vev{\Tr U^{s_1} \Tr U^{s_2} \cdots}_U
 }
where $m_{ij}^{(p)}$ label the images of $G^{(p)} (\tau_{ij})$. When
involving contractions of fermions, one replaces $G_s^{(p)}
\le(\tau_{ij} - m_{ij}^{(p)} \beta\ri)$ by $(-1)^{m_{ij}^{(p)}}
G_f^{(p)} \le(\tau_{ij} - m_{ij}^{(p)} \beta \ri)$ for the relevant
$p$'s. The powers $s_1, s_2, \cdots$ in the last factor of \genrTm\
can be found as follows. To each propagator in the diagram we assign
a direction, which can be chosen arbitrarily and similarly an
orientation can be chosen for each face. For each face $A$ in the
diagram, we have a factor $\Tr U^{s_{A}}$, with $s_A$ given by
 \eqn\idnd{
 s_A = \sum_{\p A} (\pm) m_{ij}^{(p)}, \qquad A =1,2, \cdots F
 }
where the sum $\p A$ is over the propagators bounding the face $A$
and $F$ denotes the number of faces of the diagram. In \idnd\ the
plus (minus) sign is taken if the direction of the corresponding
propagator is the same as (opposite to) that of the face.

In the low temperature phase, due to equation \factE\ one has
constraints on $m_{ij}^{(p)}$ associated with each face
 \eqn\idndR{
 s_A = \sum_{\p A} (\pm) m_{ij}^{(p)} = 0, \qquad A=1,2, \cdots F
 \ .
 }
Note that not all equations in \idndR\ are independent. The sum of
all the equations gives identically zero. One can also check that
this is the only relation between the equations, thus giving rise to
$F-1$ constraints on $m_{ij}^{(p)}$'s. For a given diagram, the
number $I$ of propagators, the number $F$ of faces and the number
$n$ of vertices\foot{Note that since we are considering the free
theory, the number of vertices coincides with the number of
operators in the correlation functions.} satisfy the relation $F+n-
I=2-2h$, where $h$ is the genus of the diagram. It then follows that
the number of independent sums over images is $K= I-(F -1) = n-1 +
2h$.

For {\it planar} diagrams, we have the number of independent sums
over images given by
 \eqn\dhjsp{
  K=n-1 \
  }
i.e. one less than the number of vertices.
 Also for any loop $L$ in a planar diagram, one
has\foot{The following equation also applies to contractible loops
in a non-planar diagram.}
 \eqn\hdks{
 \sum_{\p L} \pm m_{ij}^{(p)}=0
 }
where one sums over the image numbers associated with each
propagator that the loop contains with the relative signs given by
the relative directions of the propagators. Equation \hdks\ implies
that all propagators connecting the same two vertices should have
the same images, i.e. $m_{ij}^{(p)} = m_{ij}$ (up to a sign), which
are independent of $p$. Furthermore, this also implies that one can
write
 \eqn\akri{
 m_{ij} = m_i - m_j \ .
 }
In other words, the sums over images for each propagator reduce to
the sums over images for each operator.  We thus find that \genrTm\
becomes (for $h=0$)
 \eqn\ris{
{1 \ov N^{n-2}} \sum_{m_1, \cdots m_n=-\infty}^\infty
\prod_{i<j=1}^n \prod_{p=1}^{I_{ij}} G_s^{(p)} \le((\tau_{i}
-m_{i}\beta)- (\tau_{j} -m_{j}\beta) \ri) \ .
 }

In the above we considered contractions between different operators.
As we commented at the end of sec.2, at finite temperature
generically self-contractions do not vanish despite the normal
ordering. One can readily convinces himself using the arguments
above that all {\it planar} self-contractions reduce to those at
zero temperature and thus are canceled by normal ordering. For
example, for one-point functions, $n=1$, from \dhjsp\ there is no
sum of images. Thus the finite-temperature results are the same as
those of zero-temperature, which are zero due to normal ordering.

When the operators contain fermions, we replace $G_s$ by $G_f$ in
appropriate places and multiply \ris\ by a factor
 \eqn\ferfac{
\prod_{i<j=1}^n (-1)^{m_{ij} I^{(f)}_{ij}}
 }
where $I_{ij}^{(f)}$ is the number of fermionic propagators between
vertices $i,j$. Using \akri, we have
 \eqn\sjur{
 (-1)^{\sum_{i<j} m_{ij} I^{(f)}_{ij}} = (-1)^{\sum_{i,j} m_i I_{ij}^{(f)}} =
 (-1)^{\sum_i m_i \ep_i}
 }
 where $\ep_i =0 (1)$ if the $i$-th operator contains even (odd)
number of fermions.

Since \ris\ and \sjur\ do not depend on the specific structure of
the diagram, we conclude that to leading order in $1/N$ expansion
the full correlation function should satisfy
 \eqn\FfinrA{
 G_{\beta} (\tau_1, \cdots \tau_n) = \sum_{m_1,m_2, \cdots
 m_n=-\infty}^\infty (-1)^{m_i \ep_i} G_{0} (\tau_1-m_1 \beta, \cdots \tau_n-m_n \beta)
 \ .
 }
Note that \FfinrA\ applies also to the correlation functions in the
four dimensional theory since the harmonic expansion is independent
of the temperature.

\newsec{Including interactions}

In the sections above we have focused on the free theory limit. We
will now present arguments that \FfinrA\ remains true order by order
in the expansion over a small `t Hooft coupling $\lam$. In addition
to \onedAc\ the action also contains cubic and quartic terms which
can be written as
 \eqn\higT{
 S_{int} =N \int_0^\beta d \tau \, \le(\lam^\ha \sum_\al b_\al \LL_{3\al} + \lam
 \sum_\al d_\al \LL_{4\al}\ri)
 }
 where $\LL_{3\al}$ and $\LL_{4\al}$ are single-trace operators made from
 $\xi_a, M_a, v_a, c_a$ and their time derivatives. $b_\al$ and $d_\al$ are
numerical constants arising from the harmonic expansion.
  Again the precise form of the action
 will not be important for our discussion below. The corrections
 to free theory correlation functions can be obtained by expanding
 the exponential of \higT\ in the path integral. For example, a typical term
 will have the form
 \eqn\ryus{
  \int_{0}^\beta d \tau_{n+1} \cdots \int_{0}^\beta d
 \tau_{n+k} \, \vev{\OO_1 (\tau_1) \cdots \OO_n (\tau_n) \,
 \LL_{3\al_1} (\tau_{n+1}) \cdots \LL_{4\al_{k}}
 (\tau_{n+k})}_{\beta,0}
 }
where to avoid causing confusion we used $\vev{\cdots}_{\beta,0}$ to
denote the correlation function at zero coupling and finite
temperature. Using \FfinrA, \ryus\ can be written as\foot{Note
$\LL_{3\al}$ and $\LL_{4\al}$ also contain ghosts $c_a$ whose
contractions are temperature independent and so will not affect our
results in the last section.}
 \eqn\ryusd{
  \sum_{m_1, \cdots m_n} \int_{-\infty}^\infty d \tau_{n+1} \cdots \int_{-\infty}^\infty d
 \tau_{n+k} \, \vev{\OO_1 (\tau_1- m_1 \beta) \cdots \OO_n (\tau_n- m_n \beta) \,
 \LL_{3\al_1} (\tau_{n+1}) \cdots \LL_{4\al_{k}}
 (\tau_{n+k})}_{0,0}
 }
 where $\vev{\cdots}_{0,0}$ denotes correlation function at zero
 coupling and zero temperature and we have extended the integration ranges
for $\tau_{n+1}, \cdots \tau_{n+k}$ into $(-\infty, +\infty)$ using
the sums over the images of these variables. Equation \ryusd\ shows
that \FfinrA\ can be extended to include corrections in $\lam$.

\newsec{String theory argument and dicussions}

Equation \FfinrA, while surprising from a gauge theory point of
view, has a simple interpretation in terms of string theory dual.
Suppose the gauge theory under consideration has a string theory
dual described by some sigma-model $M$ at zero temperature and some
other sigma-model $M'$ at finite temperature. The correlation
functions in gauge theory to leading order in the $1/N$ expansion
should be mapped to sphere amplitudes of some vertex operators in
the $M$ or $M'$ theory. Equation \FfinrA\ follows immediately if we
postulate that $M'$ is identical to $M$ except that the target space
time coordinate is compactified to have a period $\beta$. To see
this, it is more transparent to write \FfinrA\ in momentum space.
Fourier transforming $\tau_i$ to $\om_i$ in \FfinrA\ we find that
 \eqn\fiss{
 G_{\beta} (\om_1, \cdots, \om_n) = G_0 (\om_1, \cdots, \om_n),
 }
with all $\om_i$ to be quantized in multiples of ${2 \pi \ov
\beta}$. Thus in momentum space to leading order in large $N$,
finite temperature correlation functions are simply obtained by
those at zero temperature by restricting to quantized momenta. From
the string theory point of view, this is the familiar inheritance
principle for tree-level amplitudes.

To use the above argument in the opposite direction, our result
suggests that in the confined phase, $M'$ should be given by $M$
with time direction periodically identified. For $\NN=4$ SYM, this
gives further support that the thermal AdS description can be
extrapolated to zero coupling\foot{See also the discussion
of~\refs{\MinW} on the extrapolation of phase diagrams
and~\refs{\FuruuchiQM} which discusses the relation between thermal
AdS and free theory correlation functions.}.

We conclude this paper by some remarks:

\item{1.} The inheritance principle~\FfinrA\ no longer holds
beyond the planar level. For non-planar diagrams, it is possible to
have images running along the non-contractible loops of the diagram.
These may be interpreted in string theory side as winding modes for
higher genus diagrams.

\item{2.} One consequence of \sunG\ is that for those correlation
functions which are independent of the 't Hooft coupling in the
large $N$ limit at zero temperature the non-renormalization theorems
remain true at finite temperature despite the fact that the
conformal and supersymmetries are broken. For $\NN=4$ SYM theory on
$S^3$, in addition to the the nonrenormalizations of two and
three-point functions of chiral operators~\refs{\MinSe} mentioned in
the Introduction, other examples include extremal correlation
functions of chiral operators~\refs{\freedman}.

\item{3.} In the high temperature (deconfined) phase, where $\Tr
U^n$ generically are non-vanishing at leading order, \FfinrA\ no
longer holds, as can be seen from the example of \twoPf. This
suggests that in the deconfined phase $M'$ should be more
complicated. In the case of $\NN=4$ SYM theory at strong coupling,
the string dual is given by an AdS Schwarzschild black
hole~\refs{\witt,\wittM}. It could also be possible that the
deconfined phase of the class of gauge theories we are considering
describe some kind of stringy black holes~\refs{\sundB,\MinW}.

 \ndt We finally note that the argument of the paper is but an example
of how the inheritance property for the sphere amplitude in an
orbifold string theory can have a non trivial realization in the
dual gauge theory. In particular it should also apply to cases where
the cycle in question is spatial rather than temporal, like the
cases discussed in~\AharonyIG.\foot{Pointed out to us by S.
Minwalla.} It would also be interesting to understand what happens
in the BTZ case.

\bigskip
\noindent{\bf Acknowledgments}

We would like to thank D.~Freedman, S.~Minwalla, S.~Shenker,
A.~Tseytlin for very useful discussions. This work is supported in
part by Alfred~P.~Sloan Foundation, DOE OJI program, and funds
provided by the Monell  Foundation and Institute for Advanced Study,
and the U.S. Department of Energy (D.O.E) under cooperative research
agreement \#DF-FC02-94ER40818.

\listrefs

\end